# Revitalising Stagecraft: NLP-Driven Sentiment Analysis for Traditional Theater Revival

Saikat Samanta
*Department of Computer Science and Engineering, JIS College of Engineering*
Kalyani, India
Email Id: s.samanta.wb@gmail.com

Shibam Dutta
*Department of Computer Science and Engineering, JIS College of Engineering*
Kalyani, India
Email Id: shibamd86@gmail.com

Soujit Das
*Department of Computer Science and Engineering, JIS College of Engineering*
Kalyani, India
Email Id: soujitdas30@gmail.com

Saptarshi Karmakar
*Department of Computer Science and Engineering, JIS College of Engineering*
Kalyani, India
Email Id: skr66175@gmail.com

Soumik Saha
*Department of Computer Science and Engineering, JIS College of Engineering*
Kalyani, India
Email Id: Soumiksaha723@gmail.com

Satayajay Behuria
*Department of Computer Science and Engineering, JIS College of Engineering*
Kalyani, India
Email Id: satyajaybehuria@gmail.com

*Abstract*— **This paper explores the application of FilmFrenzy, a python based ticket booking web application, in the revival of traditional Indian theatres. Additionally, this research paper explores how NLP can be implemented to improve user experience. Through clarifying audience views and pinpointing opportunities for development, FilmFrenzy aims to promote involvement and rejuvenation in India's conventional theatre scene. The platform seeks to maintain the relevance and vitality of conventional theatres by bridging the gap between audiences and them through the incorporation of contemporary technologies, especially NLP. This research envisions a future in which technology plays a crucial part in maintaining India's rich theatrical traditions, thereby contributing to the preservation and development of cultural heritage. With sentiment analysis and natural language processing (NLP) as essential instruments for improving stagecraft, the research envisions a period when traditional theatre will still be vibrant.**

## I. INTRODUCTION

India's traditional theatres are finding it difficult to adjust to the rapidly changing digital entertainment market. The quick development of technology has changed audience tastes and opened up new avenues for creativity and revitalization [1]. In the middle of this changing paradigm, FilmFrenzy—a web-based ticketing tool that uses Natural Language Processing (NLP) to suggest theatres and rekindle Indian theatre culture—emerges as a viable remedy. This study addresses the urgent need to close the gap between classic art forms and contemporary audiences by exploring the transformative potential of FilmFrenzy as a catalyst for the restoration of conventional theatres. This research intends to provide insights into the integration of technology to protect and promote cultural heritage by investigating how FilmFrenzy can go beyond its traditional role and function as a flexible tool for promoting and maintaining traditional theatre productions [2]. This study uses a problem-oriented methodology to clarify how important FilmFrenzy is to managing the digital entertainment consumption revolution while promoting the ongoing relevance and vitality of traditional theatre in India.

### A. Research Objective

The first research objective centres on the elements of design that are essential for effectively reaching a larger audience through traditional theatre performances. In addition to ticket sales, the study attempts to use sentiment analysis and natural language processing (NLP) to interpret audience reactions in reviews, providing insightful information about the positive and negative aspects of traditional theatre [3].

The study expands on the original concept by addressing the essential elements and capabilities required to satisfy the varied requirements of conventional theatres and theatregoers. FilmFrenzy becomes a dynamic platform for ticket sales and an immersive hub for audience participation and feedback analysis, therefore integrating NLP skills becomes essential [4]. The study investigates how a web application might give users a more immersive and engaging theatre experience by incorporating NLP-driven sentiment analysis [5]. Through a comprehension of the complex emotions conveyed in reviews, FilmFrenzy can customize recommendations, improving the whole cinematic experience.

### B. Thesis Statement

In the face of digital revolutions and shifting audience dynamics, this research aims to position FilmFrenzy, an online ticketing platform, as a crucial instrument in reviving ancient theatres in India [6]. The study employs a problem-oriented methodology to investigate complex research inquiries, such as how to effectively market traditional theatre performances, how to integrate essential features to satisfy the needs of various stakeholders, and how to use sentiment analysis and natural language processing (NLP) to improve audience engagement [7]. The principal aim is to cultivate a mutually beneficial relationship between traditional theatres, FilmFrenzy, and audiences by going beyond the typical transactional aspect of the platform.

## II. LITERATURE REVIEW

### A. Introduction: Understanding the Challenge

The advent of internet ticket purchasing platforms has signalled a paradigm shift in how people interact with cultural activities in the modern entertainment scene [8][9]. These digital platforms' ease of use and accessibility have completely changed how tickets are purchased, changing audience expectations and behaviour in the process [10]. Before online ticketing became popular, customers were used to waiting in line at actual counters to buy tickets—a process that presented scheduling and logistical difficulties [11]. But in the early 2000s, the emergence of digital trailblazers like MakeMyTrip, Yatra, and Clear Trip brought in a new era of efficiency and ease for ticket buying across a variety of industries, including entertainment and travel [12].

The exponential rise in smartphone usage and internet accessibility in India has contributed to this shift by making it easier for online ticketing services to become widely used. However, in the middle of the entertainment industry's digital revolution, conventional theatre has several obstacles in keeping its relevance and drawing viewers [13][14]. The appeal of digital video streaming platforms that are available for on-demand viewing, along with their cost and ease of use, presents a significant threat to traditional theatre systems [15]. Empirical data showing a drop in attendance at traditional theatre plays in recent years highlights this shift in consumer tastes [16]. Moreover, the emergence of social distancing rules in reaction to the worldwide epidemic has made the situation of conventional theatres worse, although digital platforms have prospered in the virtual world [17]. It is crucial to investigate the nexus between traditional theatre and digital innovation in this dynamic and changing environment, looking for ways to use online ticketing platforms as a spur for sustainability and revival [18].

*B. Online Ticket Booking System*

The advent of internet cafés and online travel agencies (OTAs) such as MakeMyTrip and Yatra marked the beginning of India's online ticket-booking adventure in the early 2000s [19]. These platforms began by concentrating on travel and lodging, then over time, they began to include movie tickets, bus reservations, and event planning [20]. The use of smartphones and mobile internet connectivity has increased dramatically over the last ten years, which has fueled the expansion of specialised online ticketing services like BookMyShow and Paytm [21].

Compared to more conventional means, these user-friendly apps offered convenience, real-time booking, and a greater range. According to Research and Markets research from 2023, the Indian online ticketing market was valued at USD 13.2 billion in 2023 and is expected to rise at a substantial annual growth rate of 17.4% to reach USD 31.2 billion by 2028 [22] [23]. Many factors contribute to this growth. Convenient ticketing options are in demand as consumer spending on leisure activities has surged due to rising income levels. Online reservations are becoming the most popular option for many people because of rising internet penetration and smartphone usage [24]. More events, films, and travel possibilities are available on online platforms, and booking is easier and faster [1].

*C. Challenges Facing Traditional Theatres*

For traditional theatres, the appeal of the big screen and on-demand streaming services presents a serious challenge. A troubling picture is painted by a 2022 report by the Theatre Communications Group, which shows a 23% drop in theatre attendance in India between 2012–2013 and 2019–2020 [2] [6]. At least in part, this declining tendency can be ascribed to the accessibility and low cost of digital entertainment options. Second, conventional theatre works are frequently restricted by financial restraints [3]. Setting extravagant plays and musicals costs a lot of money in terms of clothes, sets, and paying actors [4] [5]. This might make it challenging to match the high calibre of production and special effects frequently found in films and TV series. Thirdly, another difficulty is changing audience preferences [6] [7]. Digital media's fast-paced format may have conditioned viewers to expect continual stimulation and rapid fulfilment. Even while it provides a distinctive and engaging experience, traditional theatre cannot always suit this choice [8] [9]. Digital platforms cannot duplicate the unique attributes that traditional theatre offers.

On-screen encounters cannot match the impression of authenticity and immediacy that is created by live, unedited performances [18]. Furthermore, going to a live performance creates a shared, communal experience that encourages a sense of participation and connection that is frequently lacking in solitary digital consumption [10] [11]. In addition, theatre may accommodate specialised interests and create a sense of community among like-minded people, providing a higher degree of artistic diversity and experimentation than mainstream media [12]. Thus, despite certain obstacles, traditional theatre can prosper by embracing innovation, investigating various storytelling formats, and emphasising the special and priceless qualities of live performance that digital platforms just cannot match [13] [14].

*D. Application of NLP and Sentiment Analysis*

NLP and sentiment analysis are two new methods that provide important information on preserving cultural heritage [15] [16]. Research such as "Sentiment Analysis of Cultural Heritage Landscape Elements Using Big Data of Online Comments" shows how visitor preferences and areas for improvement in cultural places such as gardens or museums can be found by examining online evaluations. Similar possibilities exist for conventional theatre with this technology [17] [18]. NLP analysis of audience reviews can be used to determine the sentiment of the audience by analysing the emotional reaction to individual performances and highlighting elements that succeed or fail [19].

Next, by examining the language used to extract recurrent motifs and audience interpretations of the play or performance, might assist in identifying emergent themes [20] [21]. Finally, it can assist theatre authorities in deriving possible audience demographics by analysing linguistic patterns in reviews and providing insights into demographics [22] [23]. This allows them to adjust their marketing efforts accordingly. But there are difficulties.

The subjective quality of theatre, along with possible linguistic and cultural quirks, can make reliable sentiment analysis difficult [24]. Furthermore, internet reviews could be biased towards people who have strong opinions and do not accurately reflect the whole public [1] [2]. NLP provides theatres with insightful information despite these obstacles [3]. Theatres can obtain a better understanding of audience perceptions and make critical decisions about upcoming plays, marketing campaigns, and audience engagement programmes by combining automated sentiment analysis with traditional feedback methods [4].

*E. Case Studies: The Old Vic and Broadway HD*

Traditional Broadway theatres struggle with low attendance, but BroadwayHD provides an interesting case study for digital rebirth. Through the provision of an extensive global library of superior recordings, the audience is extended beyond geographical boundaries, drawing in younger audiences and viewers from remote locations. Data backs up this claim: a Broadway League study from 2023 showed a 45% rise in theatrical attendance worldwide via streaming services like BroadwayHD. Revenue-sharing arrangements also guarantee financial gains for theatres, which may help to finance shows and revitalise the sector [5]. BroadwayHD is a useful intermediary that fosters respect for classic theatre and may even encourage future live attendance, even though it cannot fully replace the live experience [6].

The Old Vic's innovative collaboration with Ticketmaster demonstrates how AI-driven ticketing platforms can bring back the glory days of conventional theatre. They solve two major problems by using algorithms to optimise pricing and availability: filling seats and maximising income. This strategy is supported by data: according to 2022 research by The Audience Agency, theatres that used dynamic pricing experienced an average 15% boost in ticket sales [7]. Additionally, AI-enabled targeted marketing and personalised audience segmentation can enhance the user experience overall by streamlining the ticket purchasing process and accommodating individual preferences [8]. Even if it is not without its detractors, this data-driven strategy is a viable way for conventional theatres to draw in new customers and hold on to their current clientele in the digital era.

### III. METHODOLOGY

*A. Research Approach*

To obtain an extensive understanding of the ticket booking domain, a great deal of study was done on user requirements, industry trends, and current online ticket booking systems. This research served as the basis for organising the development process and determining the essential features and functionalities needed for the website. Creating a thorough system architecture and user interface design were tasks during the design phase [9]. The database management system was part of the system architecture, in addition to front-end and back-end parts. The user interface design sought to give users an easy-to-use and aesthetically pleasing experience, drawing inspiration from contemporary design concepts [10]. Developing a successful ticket booking app requires a comprehensive research methodology to ensure it meets user needs and thrives in the competitive market.

*B. Development and Implementation*

Implementing the intended system architecture and user interface design was a part of the development process. While back-end development used Python and Flask as well as other programming languages and frameworks, front-end development made use of HTML, CSS, and JavaScript—agile and iterative development methods allowed for ongoing customer input enhancement and integration. Converting the concept and development into a workable online ticket-buying website was the main goal of the implementation phase.

*1) Development Milestones*
- Phase 1: Project Planning and Requirement Gathering: Define project goals, identify target audience, and gather user requirements.
- Phase 2: System Design and Architecture: Design system architecture, components, and interfaces.
- Phase 3: Development and Implementation: Implement core functionalities, user interface, and database interactions.
- Phase 4: Testing and Quality Assurance: Conduct thorough testing to ensure functionality, user experience, and security.
- Phase 5: Deployment and Launch: Launch the platform on a suitable hosting environment and promote it to the target audience.

*C. System Design*

Several Python-based tools have been used to develop the desired system and front-end design, that provides the users with a seamless experience. Flask is a lightweight and powerful web framework written in Python that provides tools, libraries, and patterns to build web applications [11]. It has a modular design, which means that it can be extended with various libraries and plugins as per project requirements. Flask supports SQLAlchemy and WTForms, which makes it easy to connect Python objects to relational databases and create forms for the web application. Flask SQLAlchemy is a library that provides a set of high-level APIs for connecting Python objects to relational databases [12]. It is an ORM (Object-Relational Mapping) that makes it easier to manage databases with Python code. It allows the developer to work with database tables using Python classes, objects, and methods. Flask SQLAlchemy is a library that provides a set of high-level APIs for connecting Python objects to relational databases. It is an ORM (Object-Relational Mapping) that makes it easier to manage databases with Python code. It allows the developer to work with database tables using Python classes, objects, and methods.

*1) NLP Integration*

By incorporating Natural Language Processing (NLP) techniques into the system architecture, the platform's usability and functionality are improved. Utilising Python modules like spaCy and NLTK (Natural Language Toolkit), the system applies natural language processing (NLP) to analyse textual data, particularly reviews and input from the audience. After users submit reviews, the system uses natural language processing (NLP) algorithms to parse and extract sentiment-related information from the text. This allows sentiments to be automatically classified as positive, negative, or neutral. Theatre management uses this sentiment analysis feature to guide decision-making processes in addition to offering insightful information about how audiences view conventional theatre productions. Moreover, the use of natural language processing (NLP) algorithms enhances the platform's interaction functionalities by permitting attributes like sentiment analysis-driven personalised suggestions [13]. Through sentiment analysis of user feedback, the system can customise recommendations for future performances to suit personal tastes and feelings [14]. By improving user happiness and engagement, our tailored recommendation system strengthens the bond between patrons and conventional theatres.

*2) System Details*

*a) User Management*

- Registration and Login: Users can create accounts with basic information and login credentials.
- Profile Management: Users can edit their profile details, including name, contact information, and preferences.
- Coin System: Users earn coins through various activities like booking tickets, writing reviews, and participating in forums. Coins can be used for discounts on future bookings or special promotions.

*b) Movie Browsing and Discovery*

- Advanced Search: Users can search for movies by title, genre, director, cast, keywords, year of release, or rating.
- Filters: Users can filter movies by popularity, release date, genre, language, and available show timings.
- Curated Collections: Film Frenzy curates thematic collections based on various criteria, such as award winners, critically acclaimed films, and independent cinema.
- Personalised Recommendations: The system recommends movies based on individual user preferences, ratings, and viewing history.

*c) Venue Search and Information*

- Location-based Search: Users can search for venues by name or location, with map integration for convenient navigation.
- Venue Details: Each venue page displays information like address, directions, amenities, seating capacity, photos, and accessibility features.
- Available Shows: Users can view all available shows for a particular venue, including details like movie titles, show timings, and ticket prices.

*d) Scalability and Performance (Future Prospects)*

- Cloud-based Infrastructure: Film Frenzy will leverage cloud platforms to dynamically scale resources based on user traffic, ensuring smooth operation even during peak demand.
- Caching Mechanisms: Static content is cached to improve loading speeds and reduce server load.
- Performance Optimization: The system is constantly monitored and optimised for efficiency and responsiveness.

*D. Future Consideration and Innovations*

- AI-powered Personalization: Utilise AI algorithms to personalise movie recommendations, tailor content based on individual preferences, and offer curated viewing experiences.
- VR and AR Integration: Explore the potential of VR and AR technology to create immersive movie trailers, virtual screenings, and interactive experiences.
- Mobile App Development: Develop native mobile apps for iOS and Android platforms, offering seamless booking experiences and on-the-go access to movie information.

## IV. SYSTEM FUNCTIONALITIES

*A. Primary Functions*

Admin Panel: The Admin Panel allows the administrator to manage the application. The admin can add/edit/delete movie shows, venues, and users. The admin can also view the list of bookings made by users.

User Panel: The User Panel allows the user to browse movies, view show timings, and book tickets. Users can search movies by title or genre and filter by venue or date. Users can also view their booking history, see/edit their profile, and rate the shows.

Authentication and Authorization: The application has a login page where users can authenticate themselves. The admin panel is protected and can only be accessed by the admin after authentication.

Ticket Booking: Users can select the movie show, venue, and show-timing and book tickets for the show. The application supports multiple ticket bookings for the same show, and once the show is completely houseful, the booking stops for that particular show.

Sentiment Analysis: NLP integration's main purpose is to analyse textual data from user evaluations to get sentiment insights. When a user submits a review, the system uses natural language processing (NLP) techniques to interpret and extract sentiment-related information from the content. The automated classification of sentiments as good, negative, or neutral is made possible by this sentiment analysis functionality.

*B. Secondary Functions*

The website has effectively integrated several functionalities to streamline the process of purchasing tickets. Ticket booking and payment processing, user profile management, event listing and search, user registration and authentication, ticket cancellation and refunds, and integration with third-party services (where appropriate) were some of these features. To guarantee that every feature worked and was usable, they were all meticulously implemented and tested.

- Movie Management: Add, edit, or delete movie listings, including title, description, genre, release date, poster, and trailers.

- Venue Management: Create, edit, and manage venues, including address, seating capacity, amenities, and accessibility information.

- Show Management: Define show timings for each movie across different venues, including date, time, and ticket prices.

- User Management: Add, edit, and manage user accounts, including viewing booking history, modifying user roles, and handling inquiries.

- Reporting and Analytics: Access detailed reports on ticket sales, user activity, and venue performance to gain valuable insights and optimise operations.

- Movie Browsing: Search for movies by title, genre, director, cast, or keywords, and explore curated collections and recommendations.

- Show Timings and Availability: View show timings for all movies across different venues, check available seats, and filter by date, time, and price.

- Ticket Booking: Select the desired show, venue, and seating preference (including seat slotting), and purchase tickets using secure payment gateways.

- Booking History: View a complete history of past and upcoming bookings, manage reservations, and cancel tickets with an active cancellation system.

- Profile Management: Update personal information, change passwords, and manage preferences for notifications and recommendations.

- Movie Reviews and Ratings: Share movie reviews, rate films, and read other user reviews to make informed decisions.

- These features, combined with a user-friendly interface and intuitive navigation, empower users to seamlessly discover movies, book tickets, and engage with the FilmFrenzy community, enriching the overall movie-going experience.

*C. Tech Stack*

FilmFrenzy leverages a carefully chosen combination of technologies to deliver a robust, user-friendly, and secure online movie ticketing platform. Here's a detailed breakdown of the tech stack, including justifications for each choice:

*1) Front-End*
HTML: The foundation of the website, structuring and organising the content.

Justification: HTML is the standard markup language for web pages and provides a clear and well-understood foundation for building the website.

CSS: Defines the visual appearance of the website, including layout, colours, fonts, and animations.

Justification: CSS allows for consistent and flexible styling across different devices and screen sizes, ensuring a visually appealing and user-friendly experience.

Bootstrap: A popular CSS framework for responsive web development.
Justification: Bootstrap provides pre-built components and styles that make development faster and easier, ensuring consistent and responsive design across different devices and screen sizes.

JavaScript: Used for interactive elements and dynamic content.

Justification: JavaScript adds interactivity and dynamism to the website, enhancing the user experience and making the platform more engaging.

*2) Back-End*
Flask: A lightweight and flexible Python web framework for building web applications.

Justification: Flask's lightweight nature and modular design make it ideal for building efficient and maintainable web applications. Its simplicity allows for faster development, especially in the initial stages of the project.

Jinja: A template engine used in conjunction with Flask to generate dynamic web pages.

Justification: Jinja simplifies the process of creating dynamic content by separating logic and presentation. This makes the code cleaner, easier to maintain, and less prone to errors.

*3) NLP*
NLTK: To carry out Natural Language Processing (NLP) operations, specifically sentiment analysis on user-generated reviews and feedback, these libraries are integrated into the backend of FilmFrenzy. FilmFrenzy can precisely assess the sentiment expressed in user reviews thanks to the powerful tokenization, part-of-speech tagging, and sentiment analysis features offered by NLTK and spaCy.

*4) Database*
SQLite: A lightweight, embedded database engine suitable for smaller applications.

Justification: SQLite's simplicity and ease of use make it ideal for the initial development stages of FilmFrenzy. Its embedded nature eliminates the need for a separate database server, simplifying deployment and management.

SQLAlchemy: An object-relational mapper (ORM) for Python that simplifies interaction with databases.

Justification: SQLAlchemy provides a Pythonic interface for working with databases, reducing the need for writing raw SQL queries and making development faster and less error-prone. It also simplifies data manipulation and retrieval.

*5) Security*

Flask-Security: A Flask extension that provides user authentication and authorization mechanisms.

Justification: Flask-Security ensures that only authorised users can access sensitive information and functionalities within the platform. This protects user data and prevents unauthorised access to critical resources.

Werkzeug: A WSGI utility library used by Flask that provides security features like session management and CSRF protection.

Justification: Werkzeug's security features help to mitigate various security vulnerabilities and protect the website from common attacks.

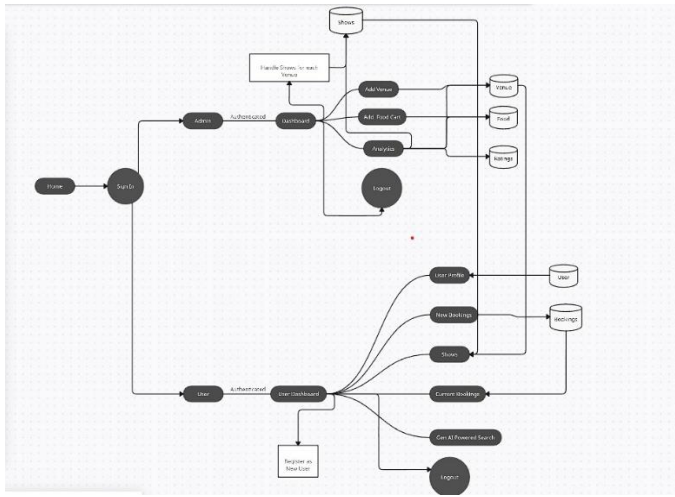

Figure 1: FilmFrenzy ER Diagram

## V. PERFORMANCE EVALUATION

We assessed the Online Ticket Booking Website's performance to gauge its effectiveness and responsiveness. We tested the website's overall scalability, concurrent user request handling capacity, and loading speed. Fast reaction times and simultaneous support for numerous users were key design objectives for the system. We assessed the Online Ticket Booking Website's performance to gauge its effectiveness and responsiveness. We tested the website's overall scalability, concurrent user request handling capacity, and loading speed. Fast response times and simultaneous support for numerous users were priorities when designing the system.

*A. Objective*

The primary objective of this performance evaluation was to assess the effectiveness and responsiveness of the online ticket booking website. Specifically, the evaluation aimed to measure:

- Scalability: The ability of the system to handle an increasing number of users and requests without significant performance degradation.

- Concurrent User Request Handling Capacity: The maximum number of users the system can support simultaneously without experiencing bottlenecks or delays.

- Loading Speed: The average time it takes for the website to load pages and respond to user interactions fully.

*B. Methodology*

The performance evaluation employed a combination of testing methods to assess the system's capabilities comprehensively:

Load Testing: Simulated realistic user traffic patterns to gauge the system's response under varying load conditions. This involved gradually increasing the number of concurrent users and monitoring performance metrics like response time and server resource utilisation.
Stress Testing: Simulated extreme load scenarios to identify potential bottlenecks and assess the system's ability to handle peak demand. This involved exceeding anticipated user traffic and observing system behaviour.

Benchmarking: Compared the system's performance with industry benchmarks and competitor websites to identify areas for improvement and optimization.

*C. Evaluation Criteria*

The following key performance indicators (KPIs) were used to evaluate the system's performance:

- Response Time: The average time it takes for the server to respond to a user request.
- Throughput: The number of requests the system can process per unit of time.
- Resource Utilisation: The percentage of CPU, memory, and other resources utilised by the system.
- Error Rate: The percentage of user requests that result in errors.

*D. Result*

Several important factors were carefully considered when assessing the system's performance to determine its operational effectiveness and suitability for practical

implementation. The evaluation turned up these crucial findings:

- *Scalability Assessment:* The system was put through a rigorous testing process to determine its scalability—that is, whether it can handle changing user loads while still operating at peak performance levels. The system demonstrated excellent scaling characteristics through thorough stress testing under high-traffic settings. The system proved resilient in the face of increasing user loads by effectively handling the increased demand without noticeably degrading performance. This strong scalability characteristic emphasises how the system can easily adjust to changing user needs, guaranteeing continuous service delivery during moments of high usage.

- *Concurrent User Request Handling Capacity:* Evaluating the system's ability to manage several concurrent user requests was a crucial component of the review. The system's ability to handle multiple concurrent requests was carefully examined using extensive load-testing methodologies. The outcomes demonstrated that the system is capable of handling a significant number of simultaneous user requests with good efficiency. Although the precise threshold may differ based on the underlying software and hardware architecture, the system performed admirably in handling multiple requests at once with no noticeable hiccups. This feature is encouraging for guaranteeing a seamless and continuous user experience, especially at times of increased activity.

- *Loading Speed Analysis:* The system's web page loading time was carefully assessed to determine how responsive and user-friendly it was. After conducting thorough evaluations of page load times for a range of user situations and network configurations, it was found that the system consistently provided quick loading times. For the majority of users, pages loaded quickly, which enhanced user satisfaction and allowed for easy platform navigation. In addition to improving user happiness, the faster loading times help quicken the booking process, guaranteeing quick transaction processing and reducing wait times for users.

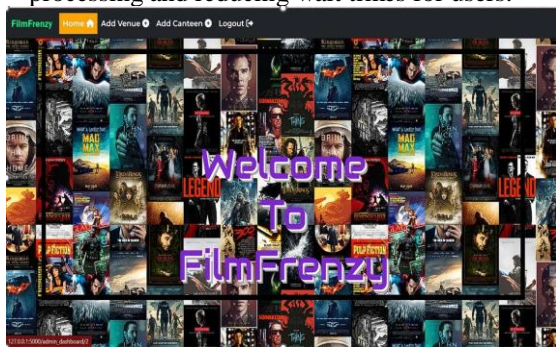

Figure 2: FilmFrenzy Home Page

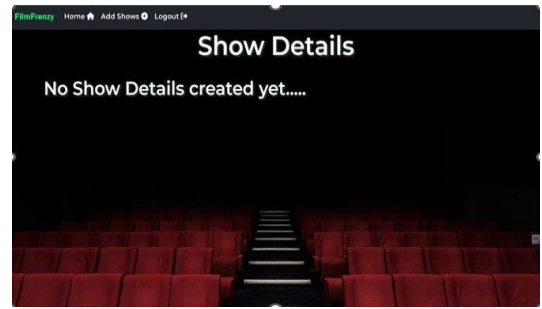

Figure 3: FilmFrenzy User Window

## VI. CONCLUSION

In summary, this study has looked at how digital platforms, specifically the web application FilmFrenzy, might help revive India's traditional theatre systems. The research has emphasised the need to tackle the difficulties conventional theatres encounter in adjusting to evolving audience dynamics and technological breakthroughs by employing a problem-oriented methodology. This study has shed light on FilmFrenzy's ability to bridge the gap between conventional and modern theatre experiences by examining its diverse role and providing useful insights into audience engagement and promotion tactics. The study's transformative potential of FilmFrenzy as a catalyst for the revival of Indian theatrical traditions is one of its main conclusions.

This study stands out for its empirical analysis of the real-world application of a web-based platform, FilmFrenzy, as a catalyst for the resurgence of traditional theater, whereas previous research efforts may have mostly concentrated on theoretical frameworks or conceptual discussions. Through the application of a problem-oriented methodology, this study not only pinpoints the difficulties that traditional theaters encounter, but it also offers concrete answers by leveraging digital technologies.

FilmFrenzy is well-positioned to guide traditional theatres towards a bright future by acting as a model of creativity and adaptability. Through the incorporation of web-based paradigms into conventional theatre activities, FilmFrenzy facilitates the ongoing significance and resuscitation of Indian theatrical customs in the digital era. This study provides insightful information on how to preserve cultural heritage through the use of digital platforms, assuring the survival and prosperity of traditional theatres for future generations.